# MAKING MAINSTREAM SYNTHESIZERS WITH CSOUND

Gleb G. Rogozinsky[1], Eugene Cherny[2, 3] and Ivan Osipenko[4]
[1]The Bonch-Bruevich Saint-Petersburg State University of Telecommunications, St. Petersburg, Russia
[2]Åbo Akademi University, Turku, Finland
[3]ITMO University, St. Petersburg, Russia
[4]Saber Interactive, St. Petersburg, Russia

For more than the past twenty years, Csound has been one of the leaders in the world of the computer music research, implementing innovative synthesis methods and making them available beyond the academic environments from which they often arise, and into the hands of musicians and sound designers throughout the world. In its present state, Csound offers an efficient environment for sound experimentation, allowing the user to work with almost any known sound synthesis or signal processing method through its vast collection of ready-made opcodes. But despite all this potential, the shared resource of Csound instruments still lacks quality reproductions of well-known synthesizers; even with its ability to generate commercial standard user interfaces and with the possibility to compile Csound instruments in such as fashion so that they can be used with no knowledge of Csound code. To fill this gap, the authors have implemented two commercial-style synthesizers as VST plug-ins using the Csound front-end 'Cabbage'. This paper describes their architecture and some of the Csound specific challenges involved in the development of fully featured synthesizers.

## 1 Wintermute

### Wintermute: the idea

In general, electronic music of the avant-garde favours atonal sounds and intervallic structures that break with convention. At the same time, contemporary commercial hardware and software synthesizers are, in their overwhelming majority, tonal. In other words, they exhibit a functional structure that is not capable of providing for the needs of modern avant-garde composers.

Of course, so-called drones *can* be created using mainstream synthesizers, but on account of their conventional design, the ability to modulate their parameters – a key aspect of ambient structures – becomes trickier, particularly when it comes to revitalizing pre-designed atonal structures. As a result, mainstream synthesizers will act more as an obstacle than a facilitating tool for composers of avant-garde electronic music.

The most popular Csound libraries and collections already offer a huge number of synthesizers, but to our minds, unfortunately, none of these instruments can compete with the most commercially successful VSTi.

We have decided to create a synth for dense dark atmospheres. We called it 'Wintermute', after the evil artificial intelligence program in William Gibson's cyberpunk novel *Neuromancer*.

**Synthesizer Structure**

The synthesizer is a harmonic drone generator, consisting of *N* voices which are uniformly distributed across the entire spectrum. Each voice consists of a white noise generator, which is passed through a volume envelope and subsequently filtered by a bandpass filter with controllable resonance. The bandwidth of each filter can be modulated so that the sound of each voice can vary between a harsh filtered noise and a sine-like signal. The central frequencies of each filter can be modulated by LFOs and envelope generators. Each voice is distributed separately in the stereo field. The voices are triggered by the gauss function. The synth idea was inspired by the 'Space Drone' synthesizer, included in the Native Instruments Reaktor library.

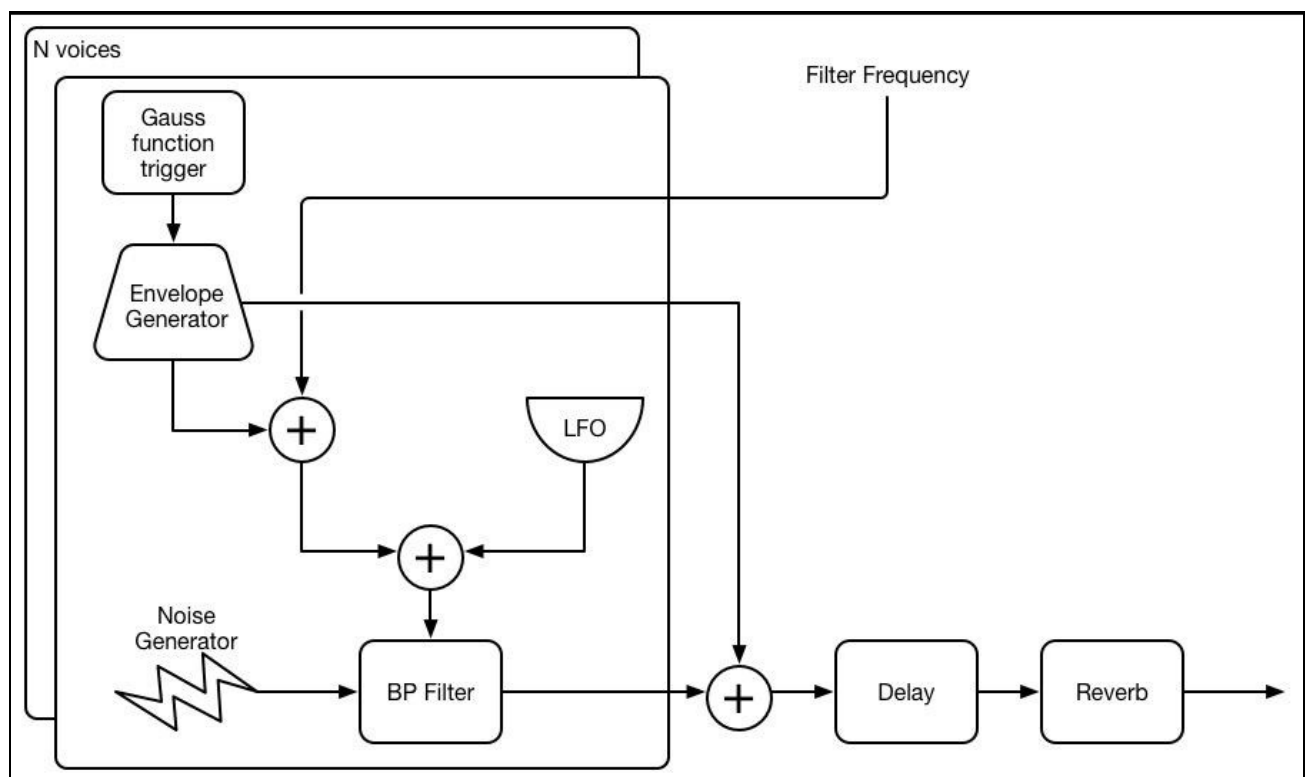

**Figure 1** The structure of the Wintermute drone generator

The FX processing block consists of a feedback delay and reverb unit based on the *reverbsc* opcode.

By default, the voice filters are tuned to the corresponding harmonics of the fundamental frequency. It produces a comb-like spectrum. The frequency spread can be changed by an additional multiplier, turning a harmonic state into an inharmonic one.

The core part of the synth is given below. The *instr 2* is a one voice generator. When it is first initialized, *instr 1* triggers N instances of *instr 2*.

```
instr 2 ; one voice
ktrig gausstrig 1, gkAvg/p6, gkDev, 0 ; random envelope re-trigger
k1 init 0
k2 rand gkAmt, .5, 0.1
k2 tonek k2, gkRHrate
k3 randomh 0.9, 1.1, gkPan*0.3
kLFO oscil 1, gkPan*k3,1,p5    ; panner lfo – each voice has separate one
kfreq=gkFun*(p4+gkOff)

label:
kLevMax rand gkAmpRand
kLevMax = 1 – kLevMax
iLev=i(k1);only at i-pass or re-init pass
k1 linsegr iLev, i(gkAtt), i(kLevMax), i(gkDec), 0, i(gkDec), 0

if ktrig>0 then
    reinit label
endif

aout resonx gaNoise*k1*k1, kfreq*2^((k1*gkMod+k2)/12), kfreq/gkRes, 2, 2
aout = aout * 0.001 * (1 + gkDamp * (p4 – p6 * .5)/p6)
aL = aout*sqrt(kLFO) ; panning with sqrt
aR = aout*sqrt(1-kLFO)
gaMixL += aL  ; sends to FX
gaMixR += aR
endin
```

## Timbral Space Analysis

In its present version, the synthesizer comes with a set of built-in presets that demonstrate the potential of the instrument in the field of sound design and procedural audio. One of its most remarkable features is its ability to morph from a dark ambient soundscape generator into a 96 (by default) voice granular-esque synth, simply by setting the appropriate parameters for its volume envelope and re-triggering rate.

1. One can easily create a quite convincing emulation of the sound of singing birds that, with appropriate spatial design of the virtual sound space, is difficult to call artificial on account of the fact that this texture is very similar to any sound in real life, and exhibits a minimum cyclical repetition of the same sounds.

> The number of voices is reduced to just a few. The envelope generators control pitch modulation and filter resonances are set to high values in order to achieve something close to tonal results. The re-triggering rate is low.

2. The next adjustment allows us to sonically simulate something akin to the boiling or bubbling of some fairly viscous liquid. Once again it is difficult to discern periodicity, which adds to the sound's authenticity.

> The number of voices is set to maximum, the re-triggering rate is set quite high and the spectrum coverage of the texture is narrowed.

3. Another remarkable example is a sound that is reminiscent of the sound of dripping liquid in a tunnel or bunker.

> First of all, reverb send should be set to to a high value. 'Drip' sounds are produced by using a fast modulation of a narrow filter.

4. Of course the synth can create many different dark ambient sounds. This was its main intended application. With the addition of some reverb and a little delay, then a raised attack and decay, you are soon in a sound world of dark ambience.

**Future Work**

The current version of the Wintermute is still at an early stage of development. The most obvious improvement needed is the creation of a user-friendly interface. The present version is lacking in this area, and development should perhaps concentrate on this first.

In addition to the improvement of the look of the synth, it is planned to diversify the modulation elements, in particular the addition of an algorithm for a looped remodulation scheme in which the filtered output is able to partially or completely act as a modulator (i.e. an LFO based on feedback).

During the sound design and preset design process, we concluded that modulation based on an LFO with a waveform selector (which would be another good addition) will work best with several oscillators which can be freely assigned to any parameter of synthesis. We also intend to implement individual envelopes or global envelopes which also can be mapped to any parameter including LFO controls.

A perfect conclusion of the work on this synthesizer would be to create a tonal, chromatic version, adapted to work with MIDI instruments and controllers. This would be particularly appropriate on account of the fact that noise based synthesizers are a rarity in the current market.

The synth was used in the performance of *BrainGame* by Tarmo Johannes at the Csound Conference. The 8-channel version of the synthesizer was used at 'Sonic Cities' (Russian Sound Art Showcase @ Ars Electronica 2015).

## 2  Shadows

**Modeling of the Access Virus sound**

The next model is, in some ways, an opposite of the first one. Having had some success with the drone synthesizer described in the first part, we decided to attempt an emulation of the famous sound of 90s era dance music - the sounds of the Access Virus, Roland JP-8000 and others.

Obviously the first element of their structure - the supersaw oscillator - is the most difficult to model. Figure… gives several original waveforms of Access Virus C. Obviously, the waves are not simply ideal sawtooths.

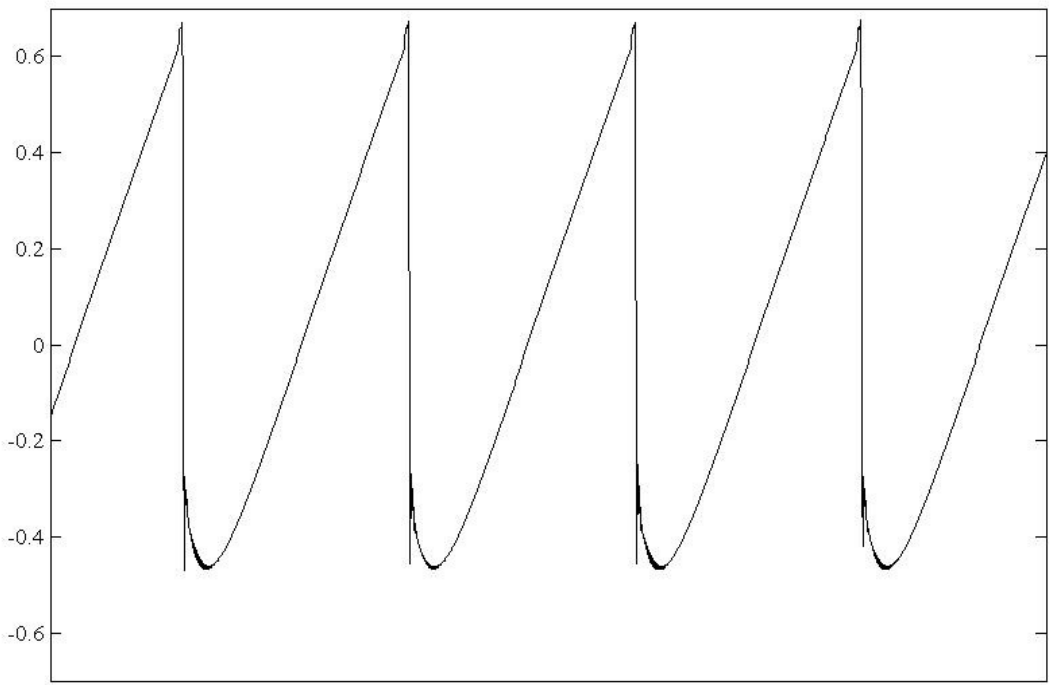

**Figure 2a**  Access Virus C sawtooth waveform

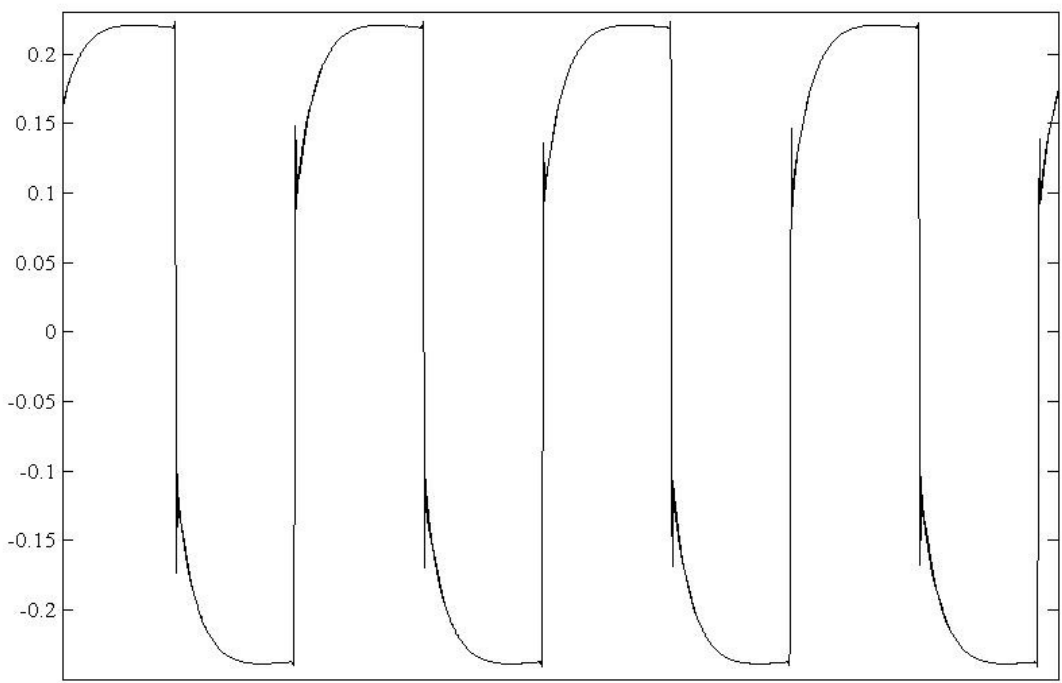

**Figure 2b**  Access Virus C pulse waveform

Therefore our main goal here was to model the oscillator waveforms as close to the reference as possible.

Gleb Rogozinsky, Eugene Cherny and Ivan Osipenko. Making Mainstream Synthesizers with Csound

The key feature of the Virus oscillator is an ability to morph between several waves (sine, triangle, saw, square and spectral waves). The *vco* and *vco2* opcodes provide that feature only for saw/triangle/ramp morphing, so we need to use a wavetable morphing solution, i.e. *tabmorphak*. To avoid aliasing we created a separate alias-free wavetable for each MIDI note. The algorithm takes a little while to load all of these tables but can be optimized by calculating function tables that can cover a range of notes as opposed to just one per note, i.e. half-octave. Below you can see a code to obtain alias-free tables for sawtooth wave (thanks to Iain McCurdy)

```
; sawtooth
iftsize    =         8192
ibaseft    =         1000
iNoteCnt           =      0
NoteLoop:
i_    ftgen    999,0,iftsize,2,0
i_    ftgen            ibaseft+iNoteCnt,0,iftsize,2,0
imaxh    =    int(sr / (2 * 440.0 * exp(log(2.0) * (iNoteCnt – 69) / 12)))
iPartCnt = 1
PartialLoop:
i_    ftgen            999,0,iftsize,-9,iPartCnt,1/iPartCnt,0
tableimix ibaseft+iNoteCnt, 0, iftsize, ibaseft+iNoteCnt,0,1, 999,0,1
loop_le iPartCnt,1,imaxh,PartialLoop
loop_le iNoteCnt,1,127,NoteLoop
```

One remarkable feature of the Access Virus or Roland JP8x is the detune control. There are several variants of the detuning. The common idea is that each time a key is pressed, several oscillators are initiated. They call voices and typically all voices share a common filter, common envelopes and the same effects, but they can be detuned between each other by some amount of cents, as well as being panned separately. Figure … shows the spectra of a heavily detuned oscillator of the T-Force Alpha synthesizer.

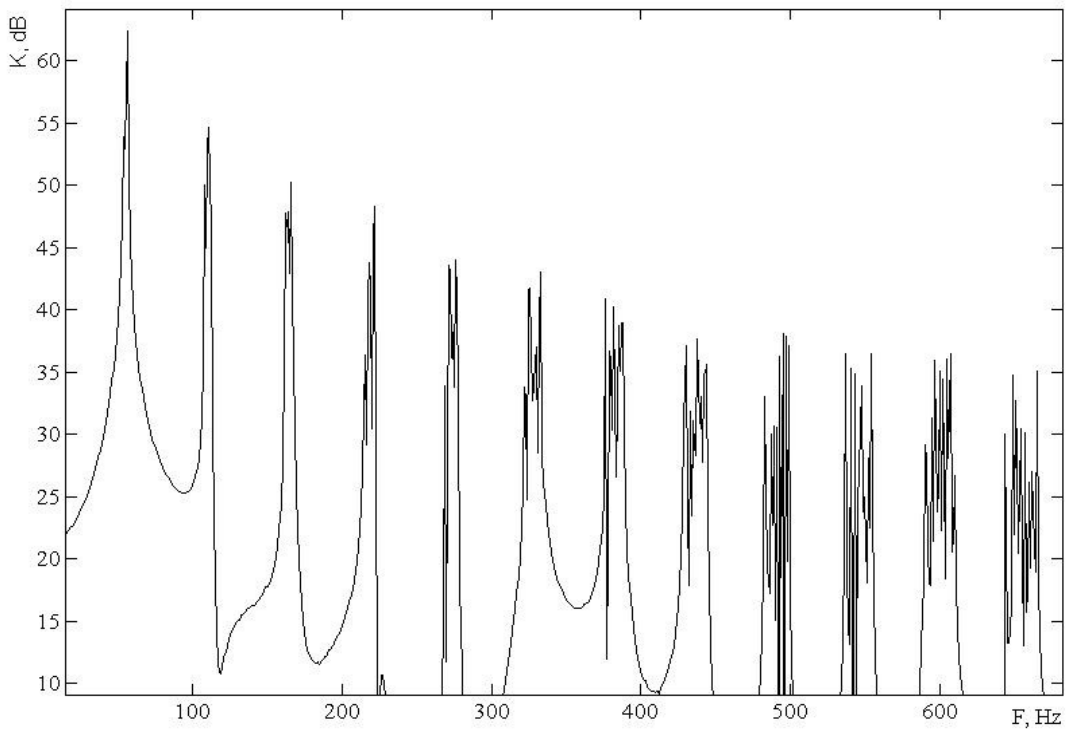

**Figure 3**   Detuned oscillator spectrum

Another feature is a panorama spread. It makes the multi-voice oscillator sound very big and wide.

Below you can see the user defined opcode *SuperOsc*. It consists of 8 instances of *MorphOsc* (which in turn holds one instance of *tabmorphak*). We used two global arrays to control the level of detune and panorama spread. As one can see, the relation between detune and spread coefficients are fixed and you can only control the amount of those effects. This is a point where, instead of having control of every tiny detail (micro-control), you can have a useful way of controlling the overall effect (macro-control).

```
opcode SuperOsc, aa, kkkkki
    setksmps 16
    kFreq, kShape, kDet, kWid, kVol, ifn xin
    kVol limit kVol, 0, 1

    gkDT[]  fillarray .0024, .019, -.019, -.0023, .0046, -.0046, .0093, -.0093
    gkP[]   fillarray .5, -.5, -.5, .5, .5, -.5, -.5, .5

    kD[]  = gkDT*kDet + 1
    kW[]  = gkP*kWid + 0.5
    kWi[] = 1 - kW

    kpw   oscil 0.01, 0.5, 1
    kpw   += 0.5

ifnSaw    = 1000 + ifn
ifnSq     = 1200 + ifn

    a1  MorphOsc kFreq*kD[0],kShape,kpw,ifnSaw,ifnSq
    a2  MorphOsc kFreq*kD[1],kShape,kpw,ifnSaw,ifnSq
    a3  MorphOsc kFreq*kD[2],kShape,kpw,ifnSaw,ifnSq
    a4  MorphOsc kFreq*kD[3],kShape,kpw,ifnSaw,ifnSq
    a5  MorphOsc kFreq*kD[4],kShape,kpw,ifnSaw,ifnSq
    a6  MorphOsc kFreq*kD[5],kShape,kpw,ifnSaw,ifnSq
    a7  MorphOsc kFreq*kD[6],kShape,kpw,ifnSaw,ifnSq
    a8  MorphOsc kFreq*kD[7],kShape,kpw,ifnSaw,ifnSq

    amixL   =   a1*kW[0]  + a2*kW[1]  + a3*kW[2]  + a4*kW[3]  + a5*kW[4]  + a6*kW[5]
+ a7*kW[6]  + a8*kW[7]
    amixR   =   a1*kWi[0] + a2*kWi[1] + a3*kWi[2] + a4*kWi[3] + a5*kWi[4] + a6*kWi[5]
+ a7*kWi[6] + a8*kWi[7]
    amixL   *=  0.125
    amixR   *=  0.125
    xout amixL*kVol, amixR*kVol
endop
```

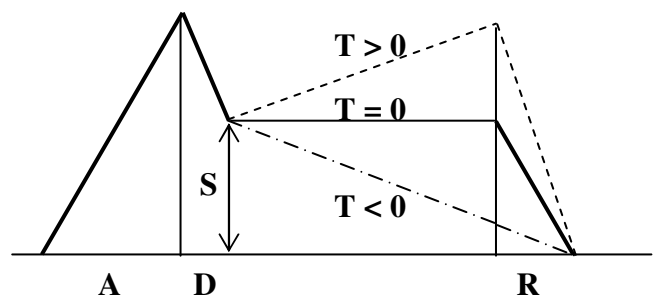

**Figure 4** ADSTR envelope

Another feature is the two ADSTR envelopes. *T* means the time the envelope takes to reach maximum (1) or minimum point. Instead of having a constant sustain value, this approach lets you have a more natural decay of level (or cut-off). The *T* value is bipolar. When it is in a neutral state (0), the envelope becomes just an ADSR envelope. A negative T value provides decay from the sustain level to 0, while positive values increase the envelope value towards 1 (maximum).

```
opcode ADSTR, kk, i
    setksmps 16
    iAmpVelocity xin

    iAmpAttack       chnget "AmpAttack"
    iAmpDecay        chnget "AmpDecay"
    iAmpSustain      chnget "AmpSustain"
    iAmpTime    chnget "AmpTime"
    iAmpRelease      chnget "AmpRelease"
    iFilAttack        chnget "FilterAttack"
    iFilDecay         chnget "FilterDecay"
    iFilSustain       chnget "FilterSustain"
    iFilTime          chnget "FilterTime"
    iFilRelease       chnget "FilterAttack"

    if iAmpTime == 0 igoto Constant
    if iAmpTime > 0 igoto Positive
    iAmpLP = 0
    iAmpTime = -iAmpTime
    goto playit
Constant:
    iAmpLP = iAmpSustain
    goto playit
Positive:
    iAmpLP = 1
playit:

    if iFilTime == 0 igoto FConstant
    if iAmpTime > 0 igoto FPositive
    iAmpLP = 0.0001
    iFilTime = -iFilTime
    goto Fplayit
FConstant:
    iFilLP = iFilSustain
    goto Fplayit
FPositive:
    iFilLP = 1
Fplayit:

    kAmpAdsr    linsegr iAmpAttack*3+0.001, iAmpDecay*3+0.0001, iAmpSustain, iAmpTime,
iAmpLP, iAmpRelease*3+0.0001, 0
    kFiltAdsr   linsegr iFilAttack*3+0.001, iFilDecay*3+0.0001, iFilSustain, iFilTime,
iFilLP, iFilRelease*3+0.0001, 0.0001

    xout kAmpAdsr*iAmpVelocity, kFiltAdsr*iAmpVelocity
endop
```

Thanks to the *Cabbage* front-end system for Csound written by Rory Walsh, we have been able to implement this synth as a VSTi, so it can be used in your favourite DAW, as well as in a stand-alone mode (using *Cabbage* or *CabbageStudio*).

The list of to-dos contains:
- Patterns. Since the synth was designed mostly for electronic dance music, it would be hugely beneficial to have an arpeggiator or at least pattern mode. It could be a non-programmable pattern, based on synced oscillator with an ability to switch between several pre-programmed function tables containing patterns.
- Phaser/Chorus. The present version of 'Shadows' lacks a Chorus/Phaser effect. The latter would be extremely useful with commercial trance arps. A chorus would be a must for Witch House music to make bass sounds even fatter.
- More precise modelling of the Access Virus original waveforms.
- A modulation matrix to provide further opportunities for more elaborate sound design.